\documentclass[11pt]{article}

\usepackage{amsmath,amsthm,amsfonts,amssymb}

\title{Geometric phase shift for detection of gravitational \\
radiation}

\author{N V Mitskievich and A I Nesterov
\\ Departamento de F\'\i sica, Universidad de Guadalajara,
\\ Guadalajara, Jalisco, M\'exico\\ E-mail: nmitskie@cencar.udg.mx,
nesterov@cencar.udg.mx}
\date{}

\begin{document}

\maketitle

\begin{abstract}
     An effect of geometrical phase shift is predicted for  a  light
beam propagating in the field of a gravitational wave. Gravitational
radiation detection experiments are proposed using  this  new  effect,
the corresponding estimates being given.
\end{abstract}

\noindent
PACS numbers: {04.80. +z }

\section{Introduction}

M. Berry (1984) has shown that a quantum system whose Hamiltonian
$H(\zeta)$ depends on some parameters $\zeta^a$ and which evolves
in time in such a way that during the evolution the state of the
system traces out a closed curve $C$ in the space of these
parameters, the wave function can get an additional geometrical phase
$\theta(C)$. This geometric phase depends on motion of the system in
the space of parameters.

Recently an effect of rotation of the polarization vector was
predicted (Chiao and Wu 1986) for a linearly polarized laser beam
travelling along a single helically wound  optical  fiber,  the  results
having  been immediately experimentally confirmed and their connection with
Berry's phase (Tomita and Chiao 1986) shown. Later it was found
(Cai et {\em al} 1990) that this Berry's phase has in  fact a classical
origin, and it arises from the intrinsic topological structure of Maxwell's
theory if the Minkowski space-time is considered as a background.  If $k$
is the wave vector of an electromagnetic  wave and ${\bf e}({\bf k})$ its
complex polarization vector, then the condition ${\bf k}^2=\omega^2({\bf
k})=$ const determines a sphere ${\rm S}^2$.  The angle $\theta({\bf
k})$ of rotation of ${\bf e}({\bf k})$, when it is parallelly
transported along a null geodesic with the tangent four-vector $k$,
is defined by

\[
\delta\theta=i {\bf e}_L({\bf k})\cdot d{\bf e}_R({\bf k}),
\]
where
\[
{\bf e}_R=2^{-1/2}({\bf e}_1+i {\bf e}_2),\qquad
{\bf e}_L=2^{-1/2}({\bf e}_1-i {\bf e}_2),\qquad
{\bf e}_L\times{\bf e}_R=i {\bf k}/\omega
\]
and ${\bf e}=e^{-\i\theta}{\bf e}_R$. The integral angle occurs to  be
$\theta=\int\delta\theta$ where  integration  is performed over the path of
${\bf k}$ on ${\rm S}^2$; thus $\theta$ has a geometric origin. When the
curve is a closed path $C$ on $S^2$, the angle $\theta(C)$ is given by
$\theta(C)=-\Omega(C)$, where $\Omega(C)$ is the solid angle of the
loop $C$ with respect to the center of the sphere.  The expression
above is essentially a flat space-time expression (see (Cai et {\em al}
1990)), however applicable also to general relativity (cf. (Bildhauer
1990)).

In the second example of the Berry's phase in optical experiments
(Pancharatnam 1956), the state space is the Poincar\'e sphere which
describes all possible polarization states of light. For this case the
direction of the light propagation is fixed and a cycle of changes in
polarization states corresponds to a closed curve on the Poincar\'e  sphere
(see e.g. (Bhandari and Manuel 1988, Simon et {\em al} 1988)). The phase
$\theta(C)$ is known as Pancharatnam phase (Pancharatnam 1956) and given by
$\theta(C)=-(1/2) \Omega(C)$.

     Here we predict a similar effect of geometrical phase shift for
light beams propagating in the field of a gravitational plane wave or
pulse of gravitational radiation:  1) cyclically from and (after
reflection) to an observer thus being closely related to the
Pancharatnam phase; 2) along a circular fiber of radius $R_0$.  We
show that for a light beam orthogonal to the direction of propagation
of a gravitational wave, this phase grows proportionally to
$L/\lambda$ where $L$ is distance between the observer and reflecting
system and $\lambda$, characteristic wavelength of the gravitational
wave packet. In the second case if $\lambda=\pi R_0$, the resonance
occurs and the phase shift grows proportionally to $m$, the number of
revolutions of light. For a pulse of gravitational radiation, the
relative phase shift is proportional to the characteristic amplitude
of the pulse in either of these two cases (for preliminary results
see (Mitskievich and Nesterov 1995)).

In this paper we use the space-time signature $(+1,-1,-1,-1)$;
Greek indices run from 0 to 3 and Latin, from 1 to 3; it is
essential to remember these notations when the integration by
parts is performed (see (4) and, in the Appendix, (9)).

\section{General results}

     A concise description of this phenomenon can be done using  the
Newman-Penrose  formalism which is applicable to propagation of light
in arbitrary media (we are interested here in the cases of (1) a
vacuum and (2) an optical fibre, the both in a gravitational field).
The real null Newman--Penrose vectors are $l=(k^0,{\bf k})$ (tangent
to the light world line) and $n ~ (l\cdot n=1)$, the complex ones,
$m$ and $\overline m$ ~ ($m\cdot\overline m=-1$). We shall consider
$m=e^{-i\theta}{\bf e}_R$ to be the circular (right) polarization
vector. When light propagates along optical fibers (not necessarily
geodesically), the Newman--Penrose description is quite essential for
differentiation along $l$, $D=\nabla_l$ being a Newman--Penrose
operator.  The corresponding equations reduce to

$$
Dl=(\epsilon+\overline{\epsilon})l-\bar{\kappa}m-\kappa\bar{m},
$$
$$
Dn=-(\epsilon+\bar{\epsilon})n+{\pi}m+\bar{\pi} \bar{m},
$$
$$
Dm=\bar{\pi}l-\kappa\bar{n}+(\epsilon-\bar{\epsilon})m.
$$
Coefficients in the right-hand side of these equations are in general
complex functions (a bar denoting complex conjugation),
$\epsilon-\bar\epsilon$ determining torsion and $\kappa$ curvature of
the space-like trajectory of light (an optical fiber); cf. (Penrose and
Rindler 1986, p. 169 ff). The change of polarization angle  generally
reads
\[
\theta=i\int{\bf e}_L\cdot D{\bf e}_R \,d\eta +\int(\bar\epsilon-\epsilon)d \eta.
\]

In a vacuum, a light beam propagates geodesically, thus
$Dl=0$, and equation of the polarization vector transport reads
$Dm=0$ (cf. [10]); these two equations yield also $Dn=0$. For a
planar optical fiber, $Dm=\bar{\pi}l-\kappa\bar{n}$. Hence in the
both cases $\bar{m}\cdot Dm$=0, thus
\begin{equation}
\theta=i\int_\Gamma{\bf e}_L\cdot D{\bf e}_R\,d\eta,
\end{equation}
the integration being performed along the light world line  $\Gamma$
canonically pa\-ra\-met\-rized by $\eta$. For the  left
polarization one has to exchange subscripts $L$ and $R$ in (1), or,
equivalently,  to change the sign in the right-hand side of this
equation.

Note that under the parallel transport along the space-like
geodesics orthogonal to the observer's world line $\gamma$,
the vector of right (left) polarization ${\bf e}_R$
(${\bf e}_L$) does not change, but it changes under the transport
along $\Gamma$ (the null line of light whose characteristics are
measured in course of the proposed experiment). This very fact
makes the existence of nontrivial phase $\theta$ essential.

If we intend to consider an experiment of detection of gravitational
radiation, it is natural to connect  the  Newman-Penrose  frame  with
the Fermi coordinates $X^\mu=\delta^\mu_0 s+\delta^\mu_i {\xi^i}u$ (see
e.g. (Manasse and Misner 1963, Misner {\em etal} 1973)). Here $s$ is proper
time along the observer's geodesic world line $\gamma$, $u$ being proper
length parametrizing the (space-like) geodesic orthogonal to it, with a
unit tangent vector ${\bf \xi}$ on $\gamma$.  In fact, ${\bf \xi}$
describes the direction in which such a space-like geodesic goes, and it
possesses only spatial components different from zero (the temporal
coordinate $X^0=T$ is directed along $\gamma$).
In Fermi coordinates, components of
the corresponding orthonormal tetrad ${e_{(\nu)}}^\mu$ parallelly
transported along the spacelike geodesic, are represented as expansions
\begin{eqnarray}
{e_{(0)}}^\mu=\delta_0^\mu-\frac{1}{2}\stackrel{0~}{R^\mu}_{ij0}X^iX^j
+\cdots,  \nonumber \\
{e_{(p)}}^\mu=\delta_p^\mu-\frac{1}{6}\stackrel{0~}{R^\mu}_{ijp}X^iX^j
+\cdots,   \nonumber
\end{eqnarray}
and the connection coefficients (Christoffel symbols) take the form
\begin{eqnarray}
\Gamma^\mu_{\nu0}=\stackrel{0~}{R^\mu}_{\nu i0}X^i +\cdots, \nonumber \\
\Gamma^\mu_{ij}= -\frac{2}{3}\stackrel{0~}{R^\mu}_{(ij)k}X^k
+\cdots.\nonumber
\end{eqnarray}
Quantities of the type of $\stackrel{0}{Q}$ are taken on $\gamma$.

The radius of convergence of series is determined by  conditions
(Manasse and Misner 1963)
\begin{equation}
u_0\ll \min \left\{ \frac{1}
{| \stackrel{0}R_{\alpha\gamma\delta\beta}|^{1/2}},\qquad
\frac{| \stackrel{0}R_{\alpha\gamma\delta\beta} |}
{| \stackrel{0}R_{\alpha\gamma\delta\beta,i}|} \right\}.
\end{equation}
The first condition $u_0 \ll | \stackrel{0}R_{\alpha\gamma\delta\beta}
|^{-1/2}$ determines the size of $V$ where the curvature has not yet caused
spatial geodesics to cross each other. The second condition defines the
domain where the curvature does not change essentially.
For instance, for gravitational waves with wavelength $\eta$ the Riemann
tensor is $\sim A\exp(i k_\mu x^\mu)/{\lambda^2}$, where $A$ is
the dimensionless amplitude. So equation (2) yields
\[
u_0\ll\min \{ \lambda/\sqrt{A},\quad  \lambda\}.
\]
Generally it is assumed $A\leq 10^{-18}$. This means that the size of $V$
is restricted by $u_0\ll\lambda$. So the application of Fermi coordinates
to the modern experiments may be very restrictive since $\lambda$ is often
supposed being in the order of 300 km. Thus for enlarging the range of
validity by a factor $1/\sqrt A$ (which is about $10^9$ in our example) it
is necesary to take into account all derivatives of the Riemann
tensor.

Covariant derivatives of the Fermi basis can be described as
(see the Appendix)
\begin{equation}
\nabla_{\partial_\lambda}{e_{(\nu)}}^\mu=
-\int_0^u{R^\mu}_{\nu\lambda p}{\xi^p}d\tau
+\frac{1}{u}\int_0^ud\tau\int_0^\tau delta^i_\lambda
{R^\mu}_{\nu ip}{\xi^p}d\tau'+O(R^2)
\end{equation}
where integration is performed along a space-like geodesic
orthogonal to the Fermi observer world line $\gamma$. This
integral formula obviously includes all derivatives of the Riemann
tensor. We shall apply equation (3) for calculation of the phase shift (1).

Using (3) we find from (1)
\[
\theta=-i\int_\Gamma  {e_L{}^\mu}
{e_R{}^\nu}{l^\lambda}d\lambda\int_0^u R_{\mu\nu\lambda p}
{\xi^p}d\tau
+i\int_\Gamma\frac{1}{u}{e_L{}^\mu}{e_R{}^\nu}d\eta
\int_0^ud\tau\int_0^\tau R_{\mu\nu ip}{l^i}
{\xi^p}d\tau'+O(R^2).
\]
Integrating by parts we obtain
\begin{eqnarray}
\theta=-i\int_\Gamma {e_L{}^\mu}{e_R{}^\nu}{l^0}d\eta
\int_0^u R_{\mu\nu 0 p}{\xi^p}d\tau
+i\int_\Gamma\frac{1}{u}{e_L{}^\mu}{e_R{}^\nu}{l^i}d\eta\int_0^u
R_{\mu\nu ip}{\xi^p}\tau d\tau+O(R^2).
\end{eqnarray}

    We consider now a plane weak gravitational wave whose
metric tensor is usually written in synchronous coordinates,
\[
{ds}^2=\eta_{\mu\nu}dx^{\mu} dx^{\nu}+{h}_{ab}dx^a dx^b,
\]
where $a$ and $b$ run from 1 to 2 while
\[
h_{a b}= h_{a b} (t-z),~  h_{2 2}= -h_{1 1} = h_{+},~ h_{1 2}=
h_{\times}.
\]
Using the definition of the Riemann tensor
\[
R_{\mu\nu\lambda\sigma}=\frac{1}{2}(h_{\nu\lambda,\mu,\sigma}+
h_{\mu\sigma,\nu\lambda}-h_{\mu\lambda,\nu,\sigma}
-h_{\nu\sigma,\mu,\lambda}),
\]
we find that in the linearized theory non-zero components of Riemann tensor
are
\begin{equation}
  R_{3ab3}=R_{0ab0}=-R_{3ab0}=\frac{1}{2}\ddot h_{ab},\qquad
R_{\mu22\nu}=-R_{\mu11\nu}, \qquad R_{\mu12\nu}=R_{\mu21\nu}
\end{equation}
where dot being derivative with respect to $t$. Now the equation (4) is
readily applicable, and two typical cases emerge:  (A) parallelly
(antiparallelly) propagating gravitational and light waves, (B) mutually
orthogonal waves.  Below we study the both cases.

\subsection{Parallelly (antiparallelly) propagating gravitational
and  light  waves}

We assume that the both waves propagate along $z$ axis. One can write
$l={l^0}(1,0,0,\pm 1)$, where $l^0=d T/d\eta$; the upper sign
corresponds to positive direction  propagation of light and the lower
sign, to negative one. Let us take
\[
e_R{}^\mu=\frac{1}{\sqrt 2}(0,1,i ,0),\qquad
e_L{}^\mu=\frac{1}{\sqrt 2}(0,1,-i ,0).
\]
Then equation (4) reduces to
\begin{equation}
\theta=-\int_0^T d \tau\int_0^{u(\tau)} R_{12 0 b} {\xi^b}d\tau'
\pm\int_0^T\frac{d \tau}{u(\tau)}\int_0^{u(\tau)}R_{123 b}
{\xi^b}\tau' d\tau'+O(R^2).
\end{equation}
Applying (5) we find $\theta=0$ that means absence of the phase shift.

\subsection{ Orthogonally propagating gravitational
and  light  waves}

Let the gravitational wave propagate along $z$ axis  and the light beam,
in the $(x,y)$ plane at an angle $\phi$ to $x$ axis. We
suppose
\begin{eqnarray}
 \xi= (0,\cos\phi,\sin\phi,0),\qquad
l={l^0}(1,\pm\cos\phi,\pm\sin\phi,0),\qquad
{l^0}={d T}/{d \eta}, \\
 {{e_R{}^\mu}}=\frac{1}{\sqrt 2}(0,\mp\sin\phi,\pm\cos\phi,i ),\qquad
{e_L{}^\mu}= \frac{1}{\sqrt 2}(0,\mp\sin\phi,\pm\cos\phi,-i ),
\end{eqnarray}
where the upper sign corresponds to propagation of light from the observer,
and the lower one, to the observer. Then the phase shift (4) is given by
\begin{eqnarray}
 \theta=\pm\int_0^T d \tau\int_0^{u(\tau)}
\bigl(R_{3120}(\cos^2\phi-\sin^2\phi) +2R_{3220}\sin\phi\cos\phi\bigr)
d\tau' + {\cal O}(R^2)
\end{eqnarray}
where we used (5). It is convenient
to rewrite this equation as \begin{eqnarray}
\theta=\mp\frac{1}{2}\int_0^Td \tau\int_0^{u(\tau)}
\bigl(\ddot h_{\times}\cos2\phi + \ddot h_{+}\sin2\phi\bigr)d\tau  +{\cal
O}(R^2).
\end{eqnarray}
Taking into account that $X^\mu=x^\mu + {\cal O}(h)$ and $h=h(t-z)$ ($h$
being $h_{\times}$ or $h_{\times}$), one can write (10) as
\begin{eqnarray}
\theta=\mp\frac{1}{2}\int_0^T\bigl(\ddot
h_{\times}(\tau)\cos2\phi + \ddot h_{+}(\tau)\sin2\phi\bigr) u(\tau) d\tau
+{\cal O}(R^2).
\end{eqnarray}
Let us define $\theta= \alpha_+\sin2\phi
+\alpha_\times\cos2\phi$, where $\alpha_+$  and $\alpha_\times$ correspond
to the two independent polarization modes of gravitational wave (see e.g.
Misner et {\em al} 1973). Then (for $\alpha_+$  or $\alpha_\times$)
\begin{equation}
\alpha=\mp\frac{1}{2}\int_0^T\ddot h(\tau)u(\tau) d\tau +{\cal O}(R^2),
\end{equation}
$h$ being $h_{+}$ or $h_{\times}$. Integration by parts with $d u={\pm}
d\tau$ yields
$$
\alpha=\frac{1}{2}\left.\Bigl(h(\tau)~ \mp~\dot
h(\tau)u(\tau)\Bigr)\right|_0^T
$$
where, as above, the upper sign corresponds to propagation of (right
polarized) light from the observer, and the lower one, to the observer. For
the left polarized light the overall sign of the right-hand side of  the
both last equations should be changed.

     Let us consider  an  experiment  when  a  circularly  polarized
electromagnetic wave  is  travelling  between  the  observer  and  a
reflecting system in the $(x,y)$ plane, the gravitational wave
propagating in the positive direction of $z$ axis. If (say, right)
polarization does not change in course of reflection, the phase shift
is
\begin{equation}
\Delta\alpha_1=\frac{1}{2}[h(2T)-h(0)-2T\dot h(T)],
\end{equation}
and if polarization changes to the left one,
\begin{equation}
\Delta\alpha_2=\frac{1}{2}[2h(T)-h(2T)-h(0)]
\end{equation}
(the effects for an  initially  left  polarized  beam  have  inverse
signs).

    For multiple reflections we have
\begin{eqnarray}
 \Delta\alpha_1(N)=\frac{1}{2}[h((N+1)T)+h(NT)-h(T)-h(0)
-2T\sum_{m=1}^{N}\dot h(mT)]   \\
\Delta\alpha_2(N)=\frac{1}{2}[h(NT)-h((N+1)T)+h(T)-h(0)],
\end{eqnarray}
where $N$ is the number of reflections.

For a plane monochromatic gravitational wave,
$h(t-z)=A\cos(\omega(t-z)+\delta)$ where $A$ is a dimensionless
amplitude.  In this case (15), (16) are rewritten as
\begin{eqnarray}
\Delta\alpha_1(N)=A(\Lambda-\sin\Lambda)
\frac{\sin(\frac{N}{2}\Lambda)\sin(\frac{N+1}{2}\Lambda+\delta)}
{\sin(\Lambda/2)},  \\
\Delta\alpha_2(N)=2A\sin(\Lambda/2)\sin(N\Lambda/2)
\cos(\frac{N+1}{2}\Lambda+\delta).
\end{eqnarray}
Here $\Lambda=2\pi L/\lambda$, $L$ being distance between the
observer and reflecting system, and $\lambda$ the gravitational
wavelength.

The average relative phase shift between the right and left polarized
light, is
\begin{eqnarray}
\sqrt{<{\Delta\theta_1}^2>}=A\left|(\Lambda-\sin\Lambda)
\frac{\sin\frac{N\Lambda}{2}}{\sin(\Lambda/2)}\right|, \\
\sqrt{<{\Delta\theta_2}^2>}=2A\left|\sin\frac{\Lambda}{2}
\sin\frac{N\Lambda}{2}\right|,
\end{eqnarray}
$A=\sqrt{A_+^2+A_\times^2}$ being the dimensionless amplitude of an
unpolarized gravitational wave, and averaging is performed with
respect to all polarizations and phases $\delta_+$ and
$\delta_\times$.

    We shall consider here no phase change in course of reflection
since this is the only way to obtain a sensible intergal effect.  If
$\Lambda\gg1$, then, for instance for $N=1$, we obtain
\begin{equation}
\sqrt{<\Delta\theta^2>}=A\Lambda.
\end{equation}
We see that effectively the
dimensionless amplitude $A$ grows by a factor $\Lambda$. When $N\gg 1$,
we find that maximum of phase shift
\begin{equation}
\sqrt{<\Delta\theta^2>}=2\pi mNA
\end{equation}
occurs for $\Lambda=2\pi m$ where $m$ is integer.  If $\Lambda\ll 1$, we
obtain
\begin{equation}
\sqrt{<\Delta\theta^2>}=
2A\left |\sin\left(\frac{N\Lambda}{2}\right)\right |,
\end{equation} and relative shift takes its maximum value,
\begin{equation}
\sqrt{<\Delta\theta^2>}=2A,
\end{equation}
when $N\Lambda=\pi$.

     Another possible experiment involves a scheme similar  to  that of
(Braginski and Menskii 1971) but with measurement of the geometric phase
shifts for right and left polarizations (and not the frequency shift as in
[15]) of light travelling along a circular fiber of radius $R_0$. Let us
consider the case when a plane gravitational wave is propagating in the
positive direction of $z$ axis, the fiber lies in $(x,y)$ plane and the
light propagates in counter-clockwise direction.  We suppose
\begin{eqnarray}
 \xi= (0,\cos\phi,\sin\phi,0),\qquad
l={l^0}(1,-\sin\phi,\cos\phi,0),\qquad {l^0}={d T}/{d \eta}, \\
 {e_R{}^\mu}=\frac{1}{\sqrt 2}(0,\cos\phi,\sin\phi,i ), \qquad
e_L{}^\mu= \frac{1}{\sqrt 2}(0,\cos\phi,\sin\phi,-i ).
\end{eqnarray}
Then the phase shift is given by
\begin{eqnarray}
 \theta=R^0\int_0^T R_{3ab0}\xi^a(\tau)\xi^b(\tau)d\tau
+{\cal O}(R^2)=
-\frac{R_0}{2}\int_0^T \ddot h_{ab}(\tau)\xi^a(\tau)\xi^b(\tau)d\tau
+{\cal O}(R^2),
\end{eqnarray}
where we used equation (5). Defining $\theta=\alpha_+ + \alpha_\times$ we
obtain

\begin{eqnarray}
\alpha_+=\frac{R_0}{2}\int_0^T\ddot h_{+}(\tau)\cos{(2\omega_0\tau)}
d\tau+O(R^2);\nonumber \\
\alpha_\times=-\frac{R_0}{2}\int_0^T\ddot h_{\times}(\tau)\sin{(2
\omega_0\tau)}d\tau + O(R^2),  \nonumber
\end{eqnarray}
where $\omega_0=2\pi/T_0$, and $T_0$ is the period of revolution of
light.  Integration by parts with $h(\tau)=A\cos(\omega\tau+\delta)$
where, as above, $A$ is  dimensionless amplitude, yields

\begin{eqnarray}
\alpha_+=A_+ \left(\frac{\omega}{2\omega_0}\right)
\left[\frac{\sin((\omega_0-\omega /2)T)\cos((\omega_0-\omega /2)T
-\delta_+)}{1-2(\omega_0/\omega)} \right. \nonumber \\
-\left. \frac{\sin((\omega_0+\omega /2)T)\cos((\omega_0+\omega /2)T-
\delta_+)}{1+2(\omega_0/\omega)} \right];  \\
\alpha_\times=A_\times\left(\frac{\omega}{2\omega_0}\right)
\left[\frac{\sin((\omega_0+\omega /2)T)\sin((\omega_0+\omega /2)T
+\delta_\times)}{1+2(\omega_0/\omega)}  \right. \nonumber\\
-\left. \frac{\sin((\omega_0-\omega /2)T)\sin((\omega_0-\omega /2)T
-\delta_\times)}{1-2(\omega_0/\omega)} \right].
\end{eqnarray}
For left polarized light, the overall sign of the right-hand side of
the both last equations should be changed. The average relative phase
shift between right and left polarized light, is

\begin{eqnarray}
\sqrt{<\Delta\theta^2>}=\frac{A~\omega}{2\sqrt{2}~\omega_0}
\left[\frac{\sin^2(\omega_0+\omega /2)T}{(1+2(\omega_0/\omega))^2}
\right. \nonumber\\
+\left.2\frac{\sin(\omega_0-\omega /2)T\sin(\omega_0+\omega /2)T\cos
(2\omega T)}{1-(2\omega_0/\omega)^2}+ \frac{\sin^2(\omega_0-
\omega /2)T}{(1-2(\omega_0/\omega))^2}\right]^{1/2}.
\end{eqnarray}

When the gravitational wavelength $\eta$ is equal to $\pi R_0$, a
resonance occurs leading to $\sqrt{<\Delta\theta^2>}=\sqrt{2}~\pi mA$
where $m$ is the number of revolutions of light. It is clear that
such a detector is effective for high frequency gravitational
radiation, the corresponding $R_0$ being around 100 km for $10^3$ Hz.
A considerably smaller size of detector could be achieved for a toroidol
winding of the fiber, this case being currently under consideration.
However measurements of the phase shift in a fiber (and in any media other
than a good vacuum) could be virtually impossible due to the random
fluctuations, so that we consider here these cases to the end of
completeness only.

\section{Discussions and conclusions}

We would like to discuss the possibility of detecting gravitational
radiation based on the proposed new effect. Let us consider the case
$\Lambda\gg 1$. From (21) we obtain $\sqrt{<\Delta\theta^2>}\approx 2\cdot
10^{-5} AL\nu$, where $\nu$ is characteristic frequency of the
gravitational radiation in Hz and $L$ is the distance in km. It is clear
that such a detector is effective for high frequency gravitational
radiation ($\nu\sim 10^4$ Hz). If, for instance, the reflecting system is
placed on the surface of moon, we have $\sqrt{<\Delta\theta^2>}\sim 8
A\nu$. For the 5-million-kilometer-long Laser Interferometer Space Antenna
(LISA), which would fly in heliocentric orbit (see Thorn 1995a,b), our
estimation of phase shift is $\sqrt{<\Delta\theta^2>}\sim 10^2 A\nu$.

Let us compare the experiment proposed here, with the standard experiments
involving resonant antennae directed to the Virgo cluster and tuned to some
3000 Hz.  In this case $A\sim 10^{-20}$ (see the corresponding  data in
(Douglas and Braginsky 1979, Thorne 1995a,b)). So the geometrical phase
detector with the base Earth -- Moon treats this radiation as if it had an
effective magnitude of some $10^{-16}$ and for LISA as if it had an
effective magnitude of some $10^{-14}$.

    For experiments using a laboratory size apparatus, $\Lambda\ll 1$, we
see that $\sqrt{<\Delta\theta^2>}\sim2A\left|\sin(N\Lambda/2)\right |$, and
the relative shift takes its maximum value when $N\Lambda=\pi$.
For LIGO/VIRGO ($L\sim 3\cdot10^5$ cm) interferometers we find that the
correspondent frequency of the gravitational wave is $\nu\sim 3\cdot
10^4/N$ .  It is known that for $L\approx10^5$cm one could expect about 300
reflections (Douglas and Braginsky 1979) which is sufficient for detection
of continuous waves with $100$ Hz $<\nu < 10^4$ Hz. The advanced LIGO
interferometers are expected to have their optimal sensitivity at $\nu\sim
100$ Hz, and rather good sensitivity all the way from $\nu \sim 10$ Hz to
$\nu\sim 500$ Hz (Thorne 1995a,b).

    If we are interested in detection of a burst of gravitational
radiation with $L\ll\lambda$  (let its maximum be at the moment of
time $NT/2$ and characteristic duration $\tau=NT$), then (15) yields
$\sqrt{<\Delta\theta^2>}\sim A$ for $N\gg1$ or $N=1$, where $A$ is
characteristic dimensionless amplitude of the pulse. Realistic
estimates for $L$ and $N$ (LIGO/VIRGO) are $L\sim 3\cdot10^5$ cm, $10< N <
300$, while the characteristic frequency is $100$ Hz $<\nu < 10^4$ Hz.

Measuring of this effect consists of a comparison of interference
patterns for the both circulary polarizations of a light beam.
Similar measurements of a phase shift between opposite senses of
circularly polarized light were performed using a nonplanar
Mach--Zehnder interferometer (Chiao et al 1988); a very clear
exposition of theoretical and experimental details concerning
observation of phase shifts due to geometrical and topological
effects, see in (Chiao 1990), naturally, without dealing with
any gravitational effects.

It is worth making a comment on the order of smallness of the
dimensionless amplitude of the gravitational wave $A$ in
connection with the limits of short gravitational
waves. This order is most invariantly attributed to the
space-time curvature which is calculated using the metric
tensor, see (5): {\bf Riem} $\sim A/\lambda^2$.
Thus when $\lambda\rightarrow 0$, $A$ should also tend to zero
if the curvature keeps its order of magnitude unchanged, and in
the limit of short gravitational wavelengths (geometric optics
in the gravitational sense) the apparent divergence
of the predicted effect at small $\lambda$'s (see equations
(17), (21)), is merely spurious:
$\sqrt{<\Delta\theta^2>}\sim\mbox{\bf Riem\,}\lambda L$.

The effect we predict in this paper, makes it in principle
possible to detect gravitational waves using not an interferometer
as a whole, but only one of its arms, since there is a
fundamental difference in propagation of the left and right
polarizations along one and the same null line of the light. One has merely
to separate the light of these different polarizations {\em after} it has
returned from its travel, then to transform the (circular) polarization of
one of the resulting beams to the opposite one, and finally to observe the
interference fringes after mixing the beams.

    We think that this effects reflects an interaction between
photon's spin and the space-time curvature, which is closely
related to the well-known Papapetrou-Mathisson effect.

\section*{Acknowledgments}
This essay was selected for an Honorable Mention by the Gravity Research
Foundation, 1994. The work was supported by CONACYT Grant No. 1626P-E9507.

\appendix
\section*{Appendix}

Here we shall obtain the formula (equation (3) in the text)
\[
\nabla_{\partial_\lambda}{e_{(\nu)}}^\mu=
-\int_0^u{R^\mu}_{\nu\lambda p}{\xi^p}d\tau
+\frac{1}{u}\int_0^ud\tau\int_0^\tau d\tau'\delta^i_\lambda
{R^\mu}_{\nu ip}{\xi^p}+{\cal O}(R^2).
\]

Let us start with the Taylor expansion for tetrad in a world tube
surrounding the world line $\gamma$ of inertial Fermi observer:
\begin{equation}
e_{(\nu)}{}^\mu= \delta^\mu_\nu
+\frac{{d\stackrel{0}e_{(\nu)}}^\mu}{d u}u
+\frac{1}{2!}\frac{{d^2\stackrel{0}e_{(\nu)}}^\mu}{d u^2}u^2
+\frac{1}{3!}\frac{{d^3\stackrel{0}e_{(\nu)}}^\mu}{d u^3}u^3+ \cdots ,
\end{equation}
$u$ being the canonical parameter along spacelike geodesics orthogonal to
$\gamma$. Using in Fermi coordinates the equation of parallel propagation
\begin{equation}
\frac{d e_{(\nu)}{}^\mu}{d u} +
\Gamma^\mu_{\sigma i}e_{(\nu)}{}^\sigma\xi^i=0,
\end{equation}
we rewrite equation (1) as
\begin{equation}
e_{(\nu)}{}^\mu= \delta^\mu_\nu
-\frac{1}{2!}\stackrel{0~}{\Gamma^{\mu}}_{\nu i,l}X^i X^l
-\frac{1}{3!}\stackrel{0~}{\Gamma^{\mu}}_{\nu i,l,p}X^i X^l X^p
+ \cdots + {\cal O}(R^2).
\end{equation}
The expansion of the connection coefficients is given by
\begin{equation}
\Gamma^\mu_{\nu\lambda}=\stackrel{0~}{\Gamma^{\mu}}_{\nu\lambda, i}X^i
+\frac{1}{2!}\stackrel{0~}{\Gamma^{\mu}}_{\nu\lambda,i,l}X^i X^l  + \cdots .
\end{equation}
Applying equations (3), (4) we obtain the following series for the spatial
covariant derivatives of tetrad \begin{eqnarray}
 {\nabla_{\partial_i}}e_{(\nu)}{}^\mu=
\stackrel{0~}{\Gamma^{\mu}}_{\nu i,k}X^k - \stackrel{0~}{\Gamma^{\mu}}_{\nu(i,k)}X^k
+\frac{1}{2!}(\stackrel{0~}{\Gamma^{\mu}}_{\nu i,k,l}
-\stackrel{0~}{\Gamma^{\mu}}_{\nu (i,k,l)}) X^k X^l  \nonumber\\
+\frac{1}{3!}(\stackrel{0~}{\Gamma^{\mu}}_{\nu i,k,l,m}
-\stackrel{0~}{\Gamma^{\mu}}_{\nu (i,k,l,m)})X^k X^l X^m
+ {\cal O}(R^2).
\end{eqnarray}
Now using the relations
\begin{eqnarray}
\frac{1}{2}\stackrel{0}{~R^\mu}{}_{\nu k i}X^k=
(\stackrel{0~}{\Gamma^{\mu}}_{\nu i,k}
-\stackrel{0~}{\Gamma^{\mu}}_{\nu(i,k)})X^k, \nonumber \\
\frac{2}{3}\stackrel{0}{~R^\mu}{}_{\nu k i,l}X^k X^l=
(\stackrel{0~}{\Gamma^{\mu}}_{\nu i,k,l}
-\stackrel{0~}{\Gamma^{\mu}}_{\nu (i,k,l)}) X^k X^l,  \nonumber\\
\frac{3}{4}\stackrel{0}{~R^\mu}{}_{\nu k i,l,m}X^k X^l X^m=
(\stackrel{0~}{\Gamma^{\mu}}_{\nu i,k,l,m}
-\stackrel{0~}{\Gamma^{\mu}}_{\nu (i,k,l,m)})X^k X^l X^m,\qquad
\mbox {etc} \nonumber
\end{eqnarray} one can write the expansion (5) as
\begin{eqnarray}
 {\nabla_{\partial_i}}e_{(\nu)}{}^\mu=
\frac{1}{2}\stackrel{0}{~R^\mu}{}_{\nu k i} X^k
+\frac{2}{3!}\stackrel{0}{~R^\mu}{}_{\nu k i,l}X^k X^l
+\frac{3}{4!}\stackrel{0}{~R^\mu}{}_{\nu k i,l}X^k X^l X^m + \cdots
\nonumber  \\
+\frac{n}{(n+1)!}\stackrel{0}{~R^\mu}{}_{\nu l_1 i,l_2, \cdots ,l_n}
X^{l_1} X^{l_2}\cdots X^{l_n}+\cdots +{\cal O}(R^2).
\end{eqnarray}
It is convenient to present this series in the form
\begin{eqnarray}
{\nabla_{\partial_i}}e_{(\nu)}{}^\mu=
\sum_{n=0}^{\infty}\frac{d^n\stackrel{0}{(R^\mu}{}_{\nu k i}
{\xi^k})}
{d u^n}\frac{u^{n+1}}{(n+2)(n!)} +{\cal O}(R^2)
\end{eqnarray}
Straightforward calculation yields the following integral representation of
equation (7):
\begin{equation}
\nabla_{\partial_i}{e_{(\nu)}}^\mu=
-\frac{1}{u}\int_0^u {R^\mu}_{\nu ik}{\xi^k} \tau d\tau
+{\cal O}(R^2).
\end{equation}
Integrating by part we find
\begin{equation}
\nabla_{\partial_i}{e_{(\nu)}}^\mu=
-\int_0^u{R^\mu}_{\nu ik}{\xi^k}d\tau
+\frac{1}{u}\int_0^ud\tau\int_0^\tau d\tau'
{R^\mu}_{\nu ik}{\xi^k}+{\cal O}(R^2).
\end{equation}

Now let us calculate the temporal covariant derivative
$\nabla_{\partial_0}{e_{(\nu)}}^\mu$. From equations (2)--(4) we easily
obtain
\begin{eqnarray}
{\nabla_{\partial_0}}e_{(\nu)}{}^\mu=
\stackrel{0~}{\Gamma^{\mu}}_{\nu 0,k}X^k
+\frac{1}{2!}(\stackrel{0~}{\Gamma^{\mu}}_{\nu 0,k,l}
-\stackrel{0~}{\Gamma^{\mu}}_{\nu k,0,l)}) X^k X^l  \nonumber\\
+\frac{1}{3!}(\stackrel{0~}{\Gamma^{\mu}}_{\nu 0,k,l,m}
-\stackrel{0~}{\Gamma^{\mu}}_{\nu k,0,l,m})X^k X^l X^m+\cdots +
+ {\cal O}(R^2),
\end{eqnarray}
where we have taken into account that
$\stackrel{0~}{\Gamma^{\mu}}_{\nu 0}=0$. Obviously,
\begin{eqnarray}
 {\nabla_{\partial_0}}e_{(\nu)}{}^\mu=
\stackrel{0}{~R^\mu}{}_{\nu k 0} X^k
+\frac{1}{2!}\stackrel{0}{~R^\mu}{}_{\nu k 0,l}X^k X^l
+\frac{1}{3!}\stackrel{0}{~R^\mu}{}_{\nu k 0,l}X^k X^l X^m + \cdots
\nonumber  \\
+\frac{1}{n!}\stackrel{0}{~R^\mu}{}_{\nu l_1 0,l_2, \cdots ,l_n}
X^{l_1} X^{l_2}\cdots X^{l_n}+\cdots +{\cal O}(R^2),
\end{eqnarray}
or
\begin{eqnarray}
{\nabla_{\partial_0}}e_{(\nu)}{}^\mu=
\sum_{n=0}^{\infty}\frac{d^n\stackrel{0}{(R^\mu}{}_{\nu k 0}
{\xi^k})}
{d u^n}\frac{u^{n+1}}{(n+1)!} +{\cal O}(R^2).
\end{eqnarray}
The correspondent integral representation reads
\begin{equation}
\nabla_{\partial_0}{e_{(\nu)}}^\mu=
-\int_0^u{R^\mu}_{\nu 0k}{\xi^k}d\tau
+{\cal O}(R^2).
\end{equation}
Finally one can combine equations (9), (13) and write
\begin{equation}
 \nabla_{\partial_\lambda}{e_{(\nu)}}^\mu=
-\int_0^u{R^\mu}_{\nu\lambda p}{\xi^p}d\tau
+\frac{1}{u}\int_0^ud\tau\int_0^\tau d\tau'\delta^i_\lambda
{R^\mu}_{\nu ip}{\xi^p}+{\cal O}(R^2).
\end{equation}

Noting that the transformation from arbitrary coordinate system to
the Fermi one takes the form $x^\mu=\Lambda^\mu{}_\nu X^\nu +O(\Gamma)$
(or $x^\mu=\xi^\mu u  +O(\Gamma)$ ) one can write Eq(14) in an arbitrary
coordinate system as
\begin{eqnarray}
\nabla_{\partial_\lambda}{e_{(\nu)}}^\mu=
-\int_0^u{R^\sigma}_{\gamma \delta\rho}(\Lambda^{-1})^\mu{}_\sigma
\Lambda^\gamma{}_\nu\Lambda^\delta{}_\lambda\xi^\rho d\tau \nonumber\\
+\frac{1}{u}\int_0^ud\tau\int_0^\tau d\tau'
{R^\sigma}_{\gamma i\rho}(\Lambda^{-1})^\mu{}_\sigma
\Lambda^\gamma{}_\nu\Lambda^i{}_\lambda\xi^\rho
+{\cal O}(R^2).
\end{eqnarray}

\newpage
~~~~~~~

\end{document}